\documentclass{article}

\PassOptionsToPackage{numbers}{natbib}
\usepackage[preprint]{neurips_2026}

\usepackage[utf8]{inputenc} 
\usepackage[T1]{fontenc}    
\usepackage{hyperref}       
\usepackage{url}            
\usepackage{booktabs}       
\usepackage{amsfonts}       
\usepackage{nicefrac}       
\usepackage{microtype}      
\usepackage{xcolor}         
\usepackage{amsmath}
\usepackage{graphicx}
\usepackage{booktabs}
\usepackage{multirow}
\usepackage[table]{xcolor}
\usepackage{placeins}
\usepackage{algorithmic}
\usepackage{algorithm}

\title{UniField: RBF-Guided Electron Density Fusion for Enhanced Molecular Representations}

\author{%
  Wei Zhang$^{\dagger}$, Kun Li$^{\dagger}$, Jiameng Chen, Wenbin Hu$^{*}$\\
  School of Computer Science\\
  Wuhan University\\
  Wuhan, China\\
  \texttt{\{zhangwei301123,likun98,jiameng.chen,hwb\}@whu.edu.cn} \\
    $^{\dagger}$These authors contributed equally. \quad
  $^{*}$Corresponding author. \\
  \And
    Yizhen Zheng \\
  Department of Data Science  and Artificial Intelligence, \\ Monash University \\
  Victoria, Australia \\
  \texttt{yizhen.zheng1@monash.edu} \\
  \And
  Jiajun Yu \\
  College of Computer Science  and Technology,\\ Zhejiang University\\
  Hangzhou, China \\
  \texttt{jiajunyu1999@gmail.com} \\
  \And
    Duanhua Cao \\
  School of Life Sciences and Technology,\\
  Tongji University\\
  Shanghai, 200092, China \\
  \texttt{caodh@tongji.edu.cn}
}

\begin{document}

\maketitle

\begin{abstract}
Current 3D geometric molecular representations predominantly focus on discrete atomic skeletons, inherently overlooking the continuous electron density (ED) field that fundamentally governs microscopic quantum behaviors. Consequently, these purely topological models suffer from critical representational blind spots, particularly in capturing long-range electron delocalization and non-covalent interactions, imposing a severe theoretical ceiling on predicting complex quantum properties. To bridge this physical gap and standardize research in electron density-enhanced molecular learning, we first construct the large-scale UniField-ED Benchmark. Comprising the QM9-ED and QMugs-ED datasets, this benchmark provides natively aligned discrete graphs and high-fidelity ED point clouds. Building upon this data infrastructure, we introduce UniField, an SE(3)-equivariant multimodal architecture that intrinsically intertwines discrete topological graphs with continuous quantum electronic environments. Extensive empirical evaluations across all three benchmarks demonstrate that UniField establishes new state-of-the-art performance. Specifically, UniField achieves a 14.8\% improvement in overall predictive performance against the leading topology-only SOTA on the ED5-OE benchmark, alongside a 37.0\% performance gain over top pure-ED models. Furthermore, on the complex drug-like dataset QMugs-ED, it yields a striking 28.2\% average precision improvement across frontier orbital properties. Alongside new SOTA results on QM9-ED, our method establishes a rigorous foundation for next-generation computational chemistry. Code and datasets are anonymously available at \url{https://anonymous.4open.science/r/UniField-ED-5B1B/}.
\end{abstract}

\section{Introduction}

With the widespread application of deep learning in areas such as materials design~\cite{xiang2025electron,qin2026msanchor} and drug discovery~\cite{yu2025collaborative,zheng2025large,icws}, the quest for more precise molecular representation \cite{li2026can,PCEvo,yu2025centrality} has become a central focus in computational chemistry~\cite{atz2021geometric,li2022deep}. Progress in this domain has been closely tied to the evolution of representational paradigms \cite{li2025drugpilot,li2025bsl}, from 1D sequence strings such as SMILES~\cite{weininger1988smiles}, InChI~\cite{heller2015inchi}, and the robust SELFIES~\cite{krenn2020selfies}, to 2D topological graphs~\cite{yu2024kernel}, and ultimately to the now dominant 3D geometric representations~\cite{aykent2025gotennet,wang2024visnet}. In particular, modeling molecules as discrete atomic graphs has been substantially advanced by networks that rigorously preserve spatial symmetries. Foundational approaches established robust SE(3)-equivariance via tensor products~\cite{thomas2018tensor,anderson2019cormorant}, inspiring both highly efficient E(n)-equivariant message passing~\cite{satorras2021egnn} and powerful recent Transformer-based architectures~\cite{liao2024equiformerv2}. However, despite their strong empirical performance, these methods still inherit a common structural bias: they represent molecules primarily as isolated atoms and interatomic relations, while overlooking the surrounding continuous quantum environment that fundamentally governs microscopic molecular behavior.

To overcome the expressive bottlenecks of purely discrete geometric representations, we incorporate a 3D electron density field. According to the Hohenberg--Kohn theorems of Density Functional Theory (DFT)~\cite{hohenberg1964inhomogeneous,kohn1965self,parr1989density}, the ground-state electron density uniquely determines the system's Hamiltonian and all ground-state properties. Consequently, an ideal molecular representation must extend beyond isolated atomic skeletons to capture electron delocalization and non-local interactions. As density-based analyses have long been essential for characterizing molecular structures, chemical bonding, and non-covalent interactions~\cite{bader1991quantum,johnson2010revealing}, integrating this microscopic quantum environment is physically rigorous and crucial for advancing molecular representation models.

Accordingly, recent studies integrate electronic information via orbital descriptors~\cite{qiao2020orbnet}, direct density prediction~\cite{jorgensen2020deepdft,jorgensen2022equivariant}, functional learning~\cite{dick2020neuralxc}, Hamiltonian learning~\cite{li2022deeph}, and density-enhanced models~\cite{xiang2025electron}. However, descriptors lack full spatial modeling, while density predictors focus on reconstruction rather than structural integration. This necessitates native fusion frameworks that preserve the geometric attributes of continuous physical fields while enabling deep cross-modal interactions between electron density points and atomic graphs.

\begin{figure}[t]
  \centering
  \includegraphics[width=\textwidth]{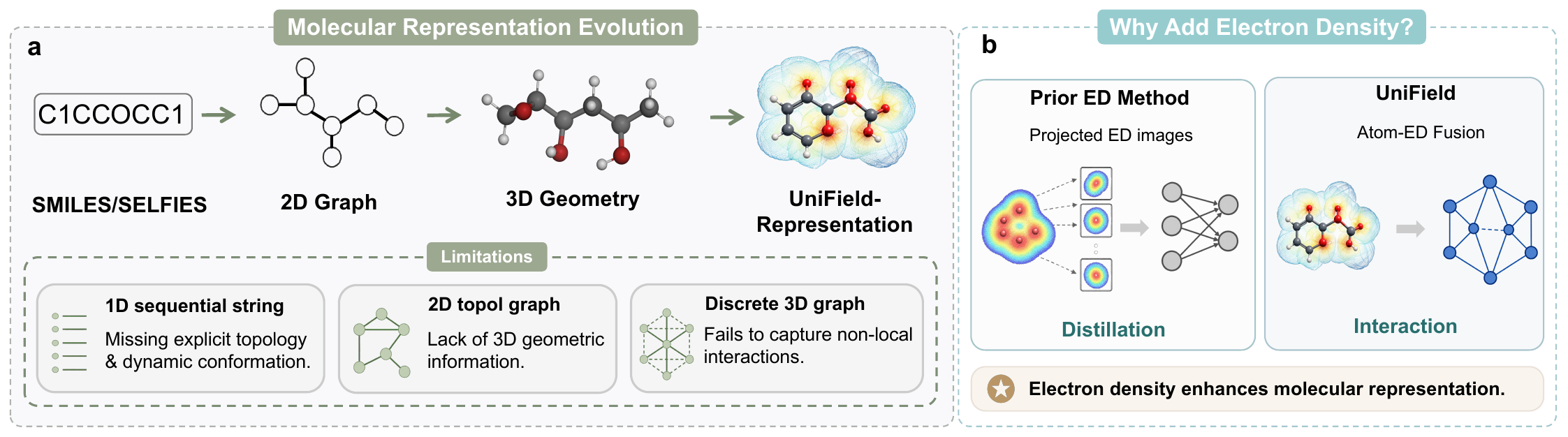}
  \vspace{-0.5cm} 
  \caption{The evolution of molecular representation paradigms and the introduction of UniField-Representation. While conventional methods have transitioned from 1D strings and 2D graphs to 3D discrete atomic skeletons, our framework represents a further paradigm shift by integrating the continuous electron density field to capture the microscopic quantum environment and non-local interactions.}
  \label{fig:motivation}
\end{figure}

To address this challenge, we propose UniField, an SE(3)-equivariant multimodal architecture for molecular representation. While preserving geometric symmetries, UniField enables direct interaction between sampled quantum fields and discrete atomic graphs at the representation level. To evaluate the proposed method, we construct the UniField-ED benchmark based on QM9~\cite{ramakrishnan2014qm9} and QMugs~\cite{isert2022qmugs}, and further assess the proposed method on both datasets as well as on the ED5-OE benchmark from EDBench~\cite{xiang2025edbench}. Experimental results demonstrate that UniField sets new SOTA results across the ED5-OE, QM9-ED, and QMugs-ED benchmarks, including a 14.8\% reduction in overall RMSE on ED5-OE and up to 29.4\% lower MAE on key electronic properties for QMugs-ED. The main contributions of this work are summarized as follows:
\begin{itemize}
 \item UniField is proposed as an SE(3)-equivariant multimodal architecture for molecular representation, incorporating high-fidelity electron density point clouds into conventional geometric molecular models. By enabling direct interactions between sampled quantum fields and discrete atomic graphs, UniField facilitates deep fusion between continuous electronic environments and molecular structures while preserving the essential geometric symmetries of molecular systems.

 \item A large-scale UniField-ED benchmark is constructed based on QM9 and QMugs to support electron-density-enhanced molecular representation learning. By providing aligned discrete atomic graphs and high-fidelity electron density point clouds, UniField-ED enables systematic study of fusion-based representations that integrate continuous quantum fields with molecular geometry in downstream prediction tasks.

\item Comprehensive evaluations on UniField-ED and ED5-OE demonstrate that incorporating electron density information improves the capture of fine-grained quantum characteristics, highlighting the potential of multimodal electron-aware representations for molecular modeling.
\end{itemize}

\section{Related Work}

Molecular systems are fundamentally governed by quantum mechanics, in which electronic structure dictates chemical and physical properties~\cite{li2025contrastive,cohen2012challenges}. Density Functional Theory (DFT) has emerged as the cornerstone of computational chemistry~\cite{burke2012perspective,van2014density} by shifting focus from many-body wavefunctions to 3D continuous electron density~\cite{parr1989density}. The Hohenberg-Kohn theorems establish ground-state density as the unique determinant of all observable properties~\cite{hohenberg1964inhomogeneous}, and alongside the Kohn-Sham equations~\cite{kohn1965self}, this paradigm provides the rigorous physical ground truth for analyzing quantum behaviors~\cite{kohn1999nobel}.

Molecular representation has evolved from 1D strings~\cite{weininger1988smiles,heller2015inchi,krenn2020selfies} and 2D graphs~\cite{gilmer2017neural,NIPS2015_f9be311e} to 3D spatial architectures~\cite{schutt2017schnet,unke2019physnet,gasteiger2020dimenetpp,liu2022spherenet}, further advanced by symmetry-preserving networks using $SE(3)$ tensor products~\cite{thomas2018tensor,anderson2019cormorant}, $E(n)$ message passing~\cite{satorras2021egnn,batzner2022nequip}, and equivariant Transformers~\cite{schutt2021painn,wang2024visnet,liao2024equiformerv2,aykent2025gotennet}. However, representing molecules solely as discrete atomic skeletons remains physically incomplete. To incorporate continuous electronic environments, recent methods integrate orbital descriptors~\cite{qiao2020orbnet}, charge densities~\cite{jorgensen2022equivariant}, functionals~\cite{dick2020neuralxc}, and Hamiltonians~\cite{li2022deeph}. Although electron density (ED) has emerged as an auxiliary representation~\cite{xiang2025electron} with specialized benchmarks~\cite{xiang2025edbench}, existing works treat ED as static descriptors, leaving native 3D fusion with discrete topologies largely unexplored.

Point clouds efficiently model 3D ED distributions without voxelization complexity. Following foundational models like PointNet/++~\cite{qi2017pointnet,qi2017pointnetplusplus}, DGCNN~\cite{wang2019dgcnn}, and Point Transformer~\cite{zhao2021pointtransformer}, recent designs including PointNeXt~\cite{qian2022pointnext}, PointVector~\cite{deng2023pointvector}, and PTv3~\cite{wu2024ptv3} have significantly enhanced scalability. While their viability for molecular property prediction is validated by EDBench~\cite{xiang2025edbench}, these vision-centric models treat ED as isolated geometric samples. Lacking intrinsic physical semantics for atomic interactions, they necessitate our proposed RBF-guided cross-modal fusion.

\section{UniField-ED Benchmark: Large-Scale Electron Density Datasets}
\label{sec:benchmark}

To address the absence of natively aligned high-fidelity electron density fields in conventional benchmarks, the UniField-ED Benchmark is introduced, comprising QM9-ED and QMugs-ED. Specifically, continuous electron densities are computed for the entire QM9 database~\cite{ramakrishnan2014qm9} via Psi4~\cite{smith2020psi4} at the B3LYP/6-31G(2df,p)+G4MP2 level. For QMugs~\cite{isert2022qmugs}, to ensure computational tractability while maintaining structural diversity, a representative subset of 50,000 drug-like conformations is curated, with densities generated at the GFN2-xTB and $\omega$B97X-D/def2-SVP levels.

\subsection{Data Construction }
\label{subsec:sampling_pipeline}

Retaining the original atomic conformations as the geometric modality $\mathcal{G}$, we discretize the corresponding continuous electron density field $\rho(\mathbf{r})$ to construct the field modality $\mathcal{P}$. To avoid the intractable $O(L^3)$ memory bottleneck of dense 3D voxel grids, we propose a two-stage physics-informed point cloud sampling strategy:

\textbf{Stage 1:} Density Thresholding Filter. We systematically discard spatial grid points where the density magnitude $e_j = \rho(\mathbf{p}_j)$ falls below a predefined threshold $\tau$. This deterministic step eliminates redundant vacuum regions and isolates structurally critical electron accumulations.

\textbf{Stage 2:} Farthest Point Sampling (FPS). To guarantee uniform spatial coverage and construct batch-compatible inputs for the point transformer, we apply FPS on the filtered valid points. The cardinality is strictly fixed to $1024$ points per molecule.

Through this pipeline, each molecule is efficiently parameterized as a standardized point cloud of shape $(1024, 4)$, encapsulating the 3D spatial coordinates $(x, y, z)$ and the scalar density magnitude $e_j$. This representation remains remarkably faithful to the microscopic electronic structure while maximizing memory efficiency.

\subsection{Dataset Composition}
\label{subsec:dataset_stats}

To rigorously evaluate the integration of continuous physical fields with discrete molecular topologies, we establish the \textbf{UniField-ED Benchmark}. As summarized in Table~\ref{tab:dataset_stats}, this benchmark bridges a critical gap in existing resources by providing large-scale, natively aligned 3D electron density point clouds and atomic graphs. The Benchmark is strategically designed to offer a hierarchical evaluation of quantum interactions: QM9-ED provides a dense, high-fidelity testbed for localized electronic states in small molecules (up to 15 atoms), whereas QMugs-ED challenges the model with multiscale structural complexities in larger, drug-like conformations (up to 80 atoms). Rather than treating property prediction as an isolated regression task, the benchmark targets are structurally partitioned into physically correlated groups. This explicit alignment with underlying quantum mechanical principles optimizes multi-task learning dynamics, enabling the network to synergistically capture shared physical priors while effectively mitigating negative transfer across distinct electronic and thermodynamic domains. Detailed benchmark construction are provided in the Appendix.

\begin{table}[htbp]
  \caption{Overview of the UniField-ED Benchmark. Both datasets provide natively aligned atomic graphs and electron density point clouds of shape (1024, 4).}
  \label{tab:dataset_stats}
  \centering
  \renewcommand{\arraystretch}{1.5} 
  \resizebox{\textwidth}{!}{
  \begin{tabular}{lccccc}
    \toprule
    \textbf{Dataset} & \textbf{Level of Theory} & \textbf{Total} & \textbf{Train/Valid/Test} & \textbf{ED Shape} & \textbf{Target Properties} \\
    \midrule
    \textbf{QM9-ED}  & B3LYP/6-31G(2df,p)+G4MP2 & 133,885 & 107,108 / 13,388 / 13,389 & $(1024, 4)$ & $\mu$, HOMO, LUMO, gap, $\alpha, \langle R^2 \rangle, \text{ZPVE}, C_v$ \\
    \midrule
    \textbf{QMugs-ED}& GFN2-xTB + $\omega$B97X-D/def2-SVP & 50,000  & 40,000 / 5,000 / 5,000  & $(1024, 4)$ & $E_{\text{form}},$ HOMO, LUMO, gap \\
    \bottomrule
  \end{tabular}
  }
\end{table}

\section{Method}
\label{sec:method}

\begin{figure}[t]
  \centering
  \includegraphics[width=\textwidth]{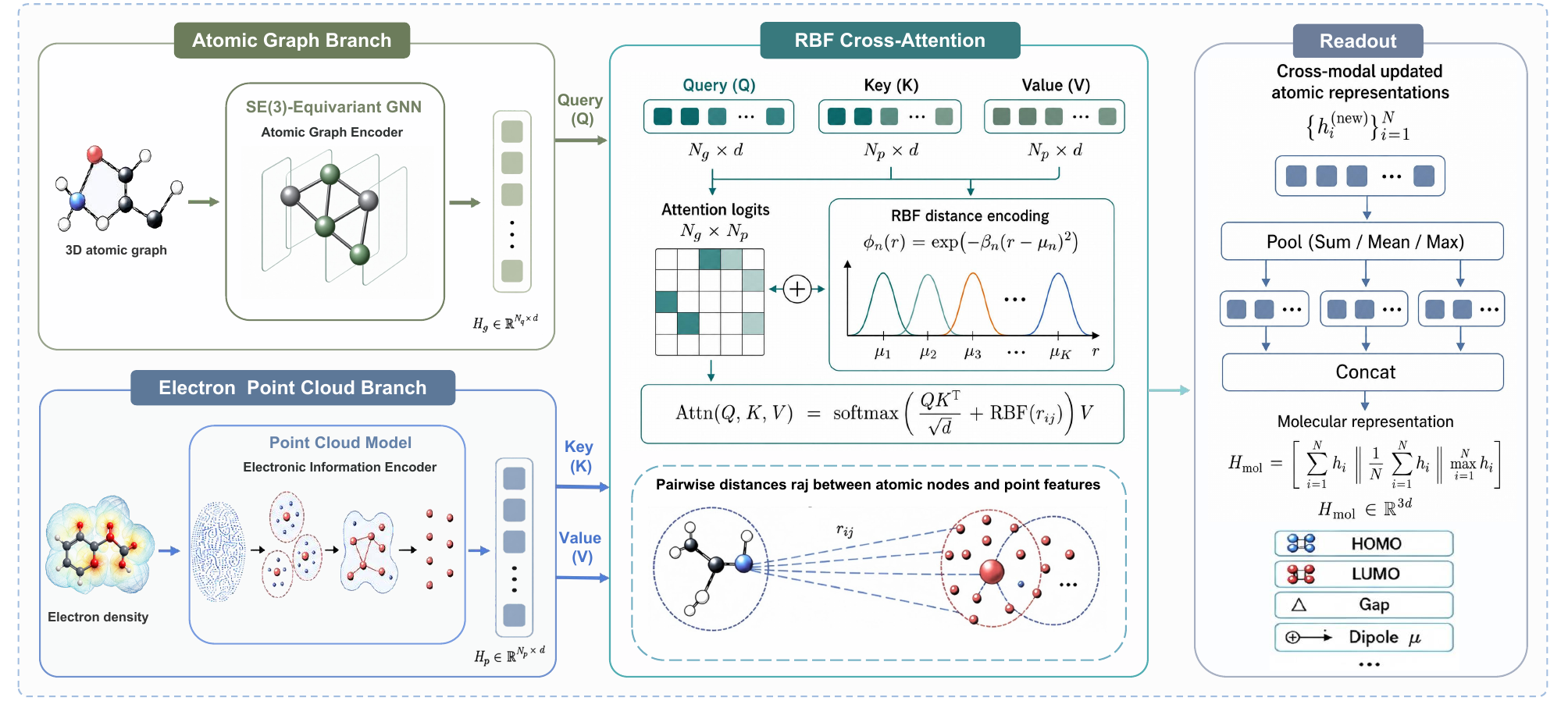}
  \vspace{-0.35cm}
  \caption{Overview of UniField. It consists of an SE(3)-Equivariant GNN for the atomic skeleton $\mathcal{G}$ and a Point Cloud Model for the electron density point cloud $\mathcal{P}$, followed by a distance-biased Electron-to-Atom Interaction module to achieve deep cross-modal feature fusion.}
  \label{fig:method}
\end{figure}

\subsection{Problem Formulation}
\label{sec:problem_formulation}

A microscopic molecular system is governed by two fundamentally intertwined physical entities: the discrete atomic topological skeleton and the surrounding electronic environment. To intrinsically integrate these entities, we formulate the molecular representation learning problem from a native 3D multimodal perspective. Let two interacting modalities define a molecular system:

\textbf{Geometric Modality ($\mathcal{G}$):} We model the discrete atomic structure as a 3D geometric graph $\mathcal{G} = (\mathcal{V}, \mathcal{E})$ with $N$ nodes. Each atom $i \in \mathcal{V}$ is defined by its atomic number $z_i \in \mathbb{N}$ and spatial coordinate $\mathbf{r}_i \in \mathbb{R}^3$, which serve as the initial node attributes and positions, respectively. Formally, this modality is defined as:

\begin{equation}
    \mathcal{G} = \left\{ (z_i, \mathbf{r}_i) \mid i = 1, 2, \dots, N \right\}
    \label{eq:modality_g}
\end{equation}

\textbf{Electron Density Field Modality ($\mathcal{P}$):} According to quantum mechanics, the electronic environment is a continuous scalar field spanning the 3D space, denoted as $\rho(\mathbf{r}): \mathbb{R}^3 \rightarrow \mathbb{R}$. To computationally process this high-dynamic-range field without falling into the curse of dimensionality typical of dense voxelization, we discretize the continuous density field into a native 3D point cloud $\mathcal{P}$. Specifically, the continuous field is sampled into $M$ discrete points:
\begin{equation}
    \mathcal{P} = \left\{ (\mathbf{p}_j, e_j) \mid j = 1, 2, \dots, M \right\}
    \label{eq:modality_p}
\end{equation}
where $\mathbf{p}_j \in \mathbb{R}^3$ and $e_j \in \mathbb{R}$ denote the spatial coordinate and scalar electron density of the $j$-th point, respectively, with $e_j$ serving as the initial 1D input feature. Ultimately, UniField aims to learn a symmetry-preserving mapping $\mathcal{F}: (\mathcal{G}, \mathcal{P}) \rightarrow \mathbf{H}_{mol} \in \mathbb{R}^{d}$ to predict downstream quantum chemical properties via independent multi-task readout heads.

\subsection{UniField Encoders: Modality-Specific Feature Extraction}
\label{sec:intra_modal}

Before the cross-modal intertwining, we extract the intra-modal physical priors from $\mathcal{G}$ and $\mathcal{P}$ independently using two modality-specific encoders.

\textbf{Atomic Skeleton Encoding.} 
We utilize an $SE(3)$-equivariant graph neural network to rigorously preserve critical spatial symmetries. This architecture leverages irreducible representations and spherical harmonics to natively model directional chemical environments. Specifically, the layer-wise atomic feature $\mathbf{h}^{\mathcal{G}}_i$ is iteratively updated via an equivariant graph attention mechanism:
\begin{equation}
    \mathbf{h}^{\mathcal{G}(l+1)}_i = \mathbf{h}^{\mathcal{G}(l)}_i + \sum_{k \in \mathcal{N}_{\mathcal{G}}(i)} \alpha_{ik} \left( \mathbf{W}_V \mathbf{h}^{\mathcal{G}(l)}_k \otimes Y(\hat{\mathbf{r}}_{ik}) \right)
\end{equation}
where $Y(\hat{\mathbf{r}}_{ik})$ denotes the spherical harmonics of the normalized directional vector between atom $i$ and its neighbor $k$, $\otimes$ represents the tensor product of irreps, and $\alpha_{ik}$ denotes the equivariant attention weights. The final layer outputs contextually enriched atomic representations $\mathbf{H}^{\mathcal{G}} = \{\mathbf{h}^{\mathcal{G}}_i\}_{i=1}^N \in \mathbb{R}^{N \times d_{\mathcal{G}}}$, which encapsulate complex geometric interactions within the molecular skeleton.

\textbf{Electron Density Field Encoding.} 
For the continuous electron density modality $\mathcal{P}$, discretized as a massive 3D point cloud, we employ a high-performance point cloud model characterized by sparse serialized self-attention~\cite{wu2024ptv3}. This architecture efficiently handles large-scale irregular point sets, extracting local density fluctuations without incurring cubic complexity. The local point feature $\mathbf{h}^{\mathcal{P}}_j$ is updated within its spatial neighborhood $\mathcal{N}_{\mathcal{P}}(j)$ as follows:
\begin{equation}
    \mathbf{h}^{\mathcal{P}(l+1)}_j = \mathbf{h}^{\mathcal{P}(l)}_j + \sum_{k \in \mathcal{N}_{\mathcal{P}}(j)} \mathrm{softmax}\left( \frac{\mathbf{q}_j^\top \mathbf{k}_k}{\sqrt{d_{\mathcal{P}}}} + \mathbf{e}_{jk} \right) \mathbf{v}_k
\end{equation}
where $\mathbf{q}_j, \mathbf{k}_k$, and $\mathbf{v}_k$ are linear projections (Query, Key, Value) of the point features, and $\mathbf{e}_{jk}$ explicitly encodes the relative spatial position bias. This hierarchical encoding yields high-level electron density representations $\mathbf{H}^{\mathcal{P}} = \{\mathbf{h}^{\mathcal{P}}_j\}_{j=1}^M \in \mathbb{R}^{M \times d_{\mathcal{P}}}$, naturally anchored at coordinates $\mathbf{p}_j$.

\subsection{UniField Fusion: Native Electron-to-Atom Interaction}
\label{sec:cross_modal}

Moving beyond conventional late-fusion concatenation, UniField introduces a native 3D cross-modal intertwining mechanism. Specifically, we design a distance-biased Electron-to-Atom Interaction layer that dynamically and rigorously anchors the continuous electronic field onto the discrete atomic skeleton, see Appendix~\ref{app:pseudocode} for the detailed algorithmic pseudocode.

\textbf{Local Spatial Alignment.} 
To ground the field information onto the atomic topology, we first establish a dynamic geometric interaction graph. For each target atom $i$ located at $\mathbf{r}_i \in \mathcal{G}$, we query the point cloud $\mathcal{P}$ to construct a local neighborhood of electron density points within a physical cutoff radius $R_c$ (such as $10$ bohr):
\begin{equation}
    \mathcal{N}(i) = \left\{ j \mid \|\mathbf{p}_j - \mathbf{r}_i\|_2 \le R_c, \ j \in \mathcal{P} \right\}
\end{equation}
This radius-based alignment restricts atomic interactions to the local quantum environment, preserving physical locality while significantly reducing computational overhead.

\textbf{Distance-Biased Cross-Attention.} 
To enable deep feature exchange from the field to the atoms, we formulate a spatially-aware cross-attention mechanism~\cite{vaswani2017attention}. Let $\mathbf{h}^{\mathcal{G}}_i \in \mathbb{R}^{d_{\mathcal{G}}}$ be the atomic feature and $\mathbf{h}^{\mathcal{P}}_j \in \mathbb{R}^{d_{\mathcal{P}}}$ be the field feature. Following the standard attention paradigm, we first linearly project these representations into Query ($\mathbf{Q}$), Key ($\mathbf{K}$), and Value ($\mathbf{V}$) spaces:
\begin{equation}
    \mathbf{q}_i = \mathbf{W}_Q \mathbf{h}^{\mathcal{G}}_i, \quad \mathbf{k}_j = \mathbf{W}_K \mathbf{h}^{\mathcal{P}}_j, \quad \mathbf{v}_j = \mathbf{W}_V \mathbf{h}^{\mathcal{P}}_j
\end{equation}
where $\mathbf{W}_Q \in \mathbb{R}^{d_{\mathcal{G}} \times d_{\mathcal{G}}}$, and $\mathbf{W}_K, \mathbf{W}_V \in \mathbb{R}^{d_{\mathcal{G}} \times d_{\mathcal{P}}}$ are learnable weight matrices.

The semantic matching score between atom $i$ and field point $j$ is initially computed via scaled dot-product: $\alpha_{ij}^{\text{base}} = (\mathbf{q}_i^\top \mathbf{k}_j) / \sqrt{d_{\mathcal{G}}}$. However, purely semantic attention lacks physical spatial awareness. To explicitly encode the isotropic spatial decay of the electron field, we introduce a continuous geometric distance bias. We calculate the Euclidean distance $d_{ij} = \|\mathbf{p}_j - \mathbf{r}_i\|_2$ and expand it using a set of Gaussian radial basis functions (RBF)~\cite{schutt2017schnet} to create a high-dimensional distance embedding $\mathbf{e}_{ij} \in \mathbb{R}^K$:
\begin{equation}
    e_{ij}^{(k)} = \exp\left( -\gamma \left(d_{ij} - \mu_k\right)^2 \right), \quad k = 1, 2, \dots, K
\end{equation}
where $K=32$ is the number of Gaussian kernels, $\mu_k$ are the uniformly spaced centers from $0$ to $R_c$, and $\gamma$ is the width coefficient.

This RBF embedding is then mapped into a scalar spatial bias via a multi-layer perceptron ($\text{MLP}_{\text{dist}}$). The spatial bias is seamlessly injected into the pre-softmax attention logits, yielding the final geometrically-biased attention score $s_{ij}$:
\begin{equation}
    s_{ij} = \frac{\mathbf{q}_i^\top \mathbf{k}_j}{\sqrt{d_{\mathcal{G}}}} + \text{MLP}_{\text{dist}}(\mathbf{e}_{ij})
\end{equation}
This addition operation  acts as a soft spatial mask, ensuring that closer electron density points exert a physically stronger influence on the atom, irrespective of the semantic feature magnitude.

\textbf{Atomic Feature Refinement.} 
The attention weights are normalized across the local field neighborhood $\mathcal{N}(i)$ using the Softmax function:
\begin{equation}
    \alpha_{ij} = \frac{\exp(s_{ij})}{\sum_{l \in \mathcal{N}(i)} \exp(s_{il})}
\end{equation}
The field messages are aggregated via a weighted sum, representing the holistic quantum influence absorbed by atom $i$:
\begin{equation}
    \mathbf{m}_i = \sum_{j \in \mathcal{N}(i)} \alpha_{ij} \mathbf{v}_j
\end{equation}
Finally, the aggregated field message $\mathbf{m}_i$ is integrated into the original atomic skeleton. To ensure optimization stability and representation robustness, the update is executed through a non-linear activation (SiLU, denoted as $\sigma$), a linear projection $\mathbf{W}_{out}$, dropout, and a residual connection followed by Layer Normalization:
\begin{equation}
    \mathbf{h}^{\mathcal{G}(\text{new})}_i = \text{LayerNorm}\left( \mathbf{h}^{\mathcal{G}}_i + \text{Dropout}\left( \mathbf{W}_{out} \sigma(\mathbf{m}_i) \right) \right)
\end{equation}
Through this meticulous mechanism, the discrete atomic graph $\mathcal{G}$ is upgraded into a unified representation $\mathcal{F}_{uni}$ that comprehensively perceives its continuous quantum electronic environment.

\subsection{Readout, Optimization, and Complexity}
\label{sec:readout_optim_complexity}

Following cross-modal fusion, the updated atomic representations 
$\mathbf{H}^{\mathcal{G}(\text{new})} = \{\mathbf{h}_i\}_{i=1}^N$ 
are aggregated into a permutation-invariant molecular representation 
$\mathbf{H}_{mol}$ through a multi-pooling readout. Specifically, we concatenate sum, mean, and max pooling over all atoms:
\begin{equation}
    \mathbf{H}_{mol} =
    \left[
    \sum_{i=1}^N \mathbf{h}_i
    \ \Bigg\| \
    \frac{1}{N}\sum_{i=1}^N \mathbf{h}_i
    \ \Bigg\| \
    \max_{i=1}^N \mathbf{h}_i
    \right].
\end{equation}
For the simultaneous prediction of $T$ quantum chemical properties, 
$\mathbf{H}_{mol}$ is fed into $T$ independent MLP readout heads,
$\hat{y}_t = \mathrm{MLP}_t(\mathbf{H}_{mol})$, which helps reduce negative transfer across heterogeneous target properties. To stabilize optimization across different physical scales, UniField is trained end-to-end with an averaged Mean Squared Error loss on Z-score normalized targets. Predictions are inverse-transformed, and Mean Absolute Error is reported as the primary metric.

We also analyze the computational complexity of UniField. Let $N_g$ and $N_p$ denote the number of atoms and electron density points, $d$ the hidden dimension, and $K \ll N_p$ the bounded number of density points within each atom's local interaction radius. The modality-specific encoders have time complexity $O(N_g d^2 + N_p d^2)$ and space complexity $O((N_g + N_p)d)$. Instead of performing global cross-attention with $O(N_g N_p d)$ complexity, UniField constructs a spatially restricted radius graph for electron-to-atom interaction, reducing the cross-modal fusion cost to $O(N_g K d)$. Therefore, the overall time and space complexities are bounded by:
\begin{equation}
    O((N_g + N_p)d^2 + N_g K d), \quad O((N_g + N_p + N_g K)d)
\end{equation}

\section{Experiment}
\label{sec:experiments}

\subsection{Experimental Setup}
\label{subsec:experimental_setup}
UniField is evaluated on three large-scale multimodal datasets. To rigorously assess generalizability and robustness on an established external testbed, the ED5-OE dataset from the recently proposed EDBench~\cite{xiang2025edbench} is first adopted. Extensive evaluations are then conducted on the UniField-ED Benchmark, detailed in Section~\ref{sec:benchmark}. These datasets provide natively aligned discrete atomic graphs and high-fidelity continuous electron density point clouds, forming a comprehensive infrastructure for evaluating multimodal field-enhanced molecular learning. To comprehensively assess model performance, UniField is benchmarked against state-of-the-art methods categorized by input modality. For the discrete geometric modality $\mathcal{G}$, representative 3D GNNs and equivariant models are selected, including SchNet~\cite{schutt2017schnet}, DimeNet++~\cite{gasteiger2020dimenetpp}, SphereNet~\cite{liu2022spherenet}, ComENet~\cite{wang2022comenet}, ViSNet~\cite{wang2024visnet}, Equiformer~\cite{liao2023equiformer}, EquiformerV2~\cite{liao2024equiformerv2}, and GotenNet~\cite{aykent2025gotennet}. For the continuous physical field modality $\mathcal{P}$, point cloud architectures including PointNeXt~\cite{qian2022pointnext} and PTv3~\cite{wu2024ptv3} are utilized.

\textbf{Implementation Details.} To prevent gradient interference, independent models are trained for each property category. Alongside the full UniField architecture, we evaluate two ablation variants: one omitting the spatial distance bias and another utilizing late fusion. Models are optimized using Adam (learning rate $10^{-4}$, weight decay $10^{-4}$) with a 1.0 gradient clipping norm and a cosine annealing scheduler. Fixed random seeds and early stopping are universally applied across all datasets to ensure reproducibility. Experiments are conducted on NVIDIA RTX 3090 and 4090 GPUs. Following convention~\cite{ramakrishnan2014qm9}, Mean Absolute Error (MAE) serves as the primary evaluation metric.

\subsection{Overall Experiment}
\label{subsec:main_results}

To rigorously demonstrate the generalizability of its native fusion mechanism, UniField is evaluated against a comprehensive suite of baselines, first on the external ED5-OE benchmark, and subsequently on our UniField-ED Benchmark. As shown in Table~\ref{tab:all_datasets_results}, UniField consistently demonstrates the advantage of integrating continuous electron density fields with discrete atomic graphs. On the external ED5-OE benchmark, UniField achieves the best overall RMSE of $161.220$ meV, corresponding to a $14.81\%$ error reduction over the strongest geometry-only baseline, EquiformerV2, and a $37.02\%$ reduction over the strongest ED-only baseline, PTv3. This confirms the importance of jointly modeling the atomic skeleton and the surrounding electronic field: pure point cloud models lack molecular topological anchoring, while geometry-only GNNs miss continuous non-local electronic information. On QM9-ED, UniField obtains competitive or best performance on strictly electronic properties, including dipole moment and HOMO--LUMO gap. However, the performance gains on this dataset are less dominant compared to larger benchmarks. We attribute this to the relatively simple and rigid nature of small QM9 molecules, where macroscopic thermodynamic properties (e.g., $\alpha$, $R^2$, and $C_v$) are already exceptionally well-defined by the discrete nuclear skeleton alone. For such targets, forcing the network to co-optimize high-frequency local density fluctuations introduces slight optimization noise, allowing geometry-only baselines to retain a marginal advantage. Nonetheless, on the more complex drug-like QMugs-ED dataset, UniField decisively achieves the best results on HOMO, LUMO, and HOMO--LUMO gap prediction, with an average $28.2\%$ error reduction over the strongest geometry-only baseline across these frontier orbital properties. These results strongly suggest that the benefit of electron-density-enhanced fusion becomes increasingly pronounced as molecular size and electronic complexity scale up, highlighting UniField as a highly effective and scalable representation framework for multimodal molecular learning.

\begin{table*}[t]
  \caption{Comprehensive performance comparison (MAE) across ED5-OE, QM9-ED and QMugs-ED. The best results are \textbf{bolded} and the second-best are \underline{underlined}.}
  \label{tab:all_datasets_results}
  \centering
  \renewcommand{\arraystretch}{1.15}
  \resizebox{\textwidth}{!}{
  \begin{tabular}{l ccccccc ccc}
    \toprule
    \textbf{Properties/Methods} & \textbf{SchNet} & \textbf{DimeNet++} & \textbf{SphereNet} & \textbf{ComENet} & \textbf{ViSNet} & \textbf{EqV2} & \textbf{GotenNet} & \textbf{PointNeXt} & \textbf{PTv3} & \textbf{UniField} \\
    \midrule
    \multicolumn{11}{l}{\cellcolor{gray!10}\textit{\textbf{Dataset: ED5-OE}}} \\
    HOMO-2 (meV) & 225.803 & 297.210 & 267.987 & 157.029 & 166.323 & \underline{125.080} & 162.490 & 488.987 & 173.926 & \textbf{109.434} \\
    HOMO-1 (meV) & 221.697 & 255.152 & 234.592 & 146.272 & 151.975 & \underline{114.115} & 150.209 & 482.351 & 173.610 & \textbf{96.427} \\
    HOMO-0 (meV) & 254.061 & 249.480 & 229.586 & 146.729 & 141.016 & \underline{112.254} & 141.640 & 549.545 & 183.063 & \textbf{94.017} \\
    LUMO+0 (meV) & 309.125 & 390.635 & 321.175 & 158.423 & 152.041 & \underline{120.985} & 158.828 & 934.591 & 207.772 & \textbf{102.821} \\
    LUMO+1 (meV) & 276.189 & 402.032 & 316.529 & 157.442 & 171.469 & \underline{138.206} & 178.566 & 874.623 & 196.935 & \textbf{115.800} \\
    LUMO+2 (meV) & 257.099 & 388.750 & 314.163 & 166.013 & 188.736 & \underline{143.295} & 185.308 & 956.061 & 189.685 & \textbf{122.128} \\
    LUMO+3 (meV) & 246.324 & 388.232 & 296.436 & 172.474 & 189.682 & \underline{147.825} & 182.991 & 942.168 & 184.985 & \textbf{123.512} \\
    \textbf{RMSE} & 337.733 & 442.322 & 374.793 & 225.174 & 234.046 & \underline{189.245} & 234.877 & 913.794 & 255.982 & \textbf{161.220} \\
    \midrule
    \multicolumn{11}{l}{\cellcolor{gray!10}\textit{\textbf{Dataset: QM9-ED}}} \\
    $\mu$ (mD) & 202.887 & 67.049 & 170.569 & 87.997 & \underline{37.361} & 73.970 & 38.003 & 1132.430 & 390.011 & \textbf{36.452} \\
    HOMO (meV) & 112.640 & 43.416 & 57.717 & 54.839 & \textbf{37.304} & 56.528 & 39.179 & 434.525 & 112.972 & \underline{37.563} \\
    LUMO (meV) & 131.706 & 49.619 & 69.076 & 64.528 & \textbf{42.291} & 73.441 & 44.647 & 1026.206 & 111.704 & \underline{42.472} \\
    $\Delta\varepsilon$ (meV) & 166.050 & 64.600 & 89.013 & 80.725 & \underline{55.366} & 88.688 & 57.926 & 1062.960 & 204.680 & \textbf{55.154} \\
    $\alpha$ (mBohr$^3$) & 517.392 & 264.249 & 344.020 & 310.898 & \underline{236.472} & 294.902 & \textbf{203.269} & 6115.220 & 921.904 & 256.416 \\
    $R^2$ (mBohr$^2$) & 13956.431 & 5601.662 & 8389.077 & 5813.466 & \underline{4511.053} & 5804.217 & \textbf{3103.456} & 191630.000 & 26046.619 & 4969.021 \\
    ZPVE (meV) & 12.299 & 11.203 & 14.635 & 10.902 & 9.745 & 9.729 & \textbf{6.841} & 720.134 & 146.706 & \underline{8.974} \\
    $C_v$ (mcal/molK) & 159.748 & 102.226 & 131.004 & 116.596 & \underline{100.879} & 117.501 & \textbf{81.878} & 3078.654 & 332.252 & 107.778 \\
    \midrule
    \multicolumn{11}{l}{\cellcolor{gray!10}\textit{\textbf{Dataset: QMugs-ED}}} \\
    HOMO (meV) & 204.387 & 133.019 & 213.306 & 109.513 & 137.168 & \underline{86.019} & 105.021 & 317.708 & 216.432 & \textbf{62.492} \\
    LUMO (meV) & 237.571 & 143.874 & 267.057 & 109.558 & 138.027 & \underline{88.890} & 117.128 & 344.008 & 240.792 & \textbf{64.118} \\
    $\Delta\varepsilon$ (meV) & 295.143 & 198.940 & 339.014 & 147.039 & 189.555 & \underline{119.872} & 152.264 & 427.231 & 295.404 & \textbf{84.592} \\
    $\Delta E_f$ (meV) & 1118.318 & 5353.438 & 21643.865 & 2566.645 & 4703.976 & \textbf{455.922} & 2661.892 & 6665.229 & 7832.953 & \underline{511.135} \\
    \bottomrule
  \end{tabular}
  } 
\end{table*}

\subsection{Ablation Study}
\label{subsec:ablation}

Spatially aligned fusion and RBF-guided distance bias are the two core components of the UniField framework.

\noindent\textbf{\textit{Q1: Is spatially-grounded fusion superior to naive late fusion?}} Comparing our model against a late-fusion variant (Tables~\ref{tab:ablation_combined} and \ref{tab:ablation_qmugsed}) reveals that naive fusion  degrades performance. Notably, QMugs-ED formation energy errors ($\Delta E_f$) spike to $3357.337$ meV, confirming that flattening 3D density into a global vector destroys critical spatial correspondences. Conversely, spatially-grounded fusion tightly anchors density fluctuations to the local nuclear topology, preventing structural collapse.

\noindent\textbf{\textit{Q2: How does the RBF-guided distance bias contribute?}} Removing this bias uniformly degrades frontier orbital predictions such as HOMO, LUMO, and $\Delta\varepsilon$. across all benchmarks. By explicitly encoding the electron field's physical decay, the RBF bias prevents feature over-smoothing and effectively captures highly delocalized states. This physically grounded injection proves essential for minimizing orbital errors in both small molecules and complex conformations.

\noindent\textbf{\textit{Q3: Does the ED modality uniformly improve all properties?}} QM9-ED exposes a fundamental representation trade-off: UniField dominates strictly electronic targets ($\mu$, energy gaps), while the geometry-only baseline retains a marginal advantage on macroscopic thermodynamic properties such as $\alpha$ and $R^2$. This implies that co-optimizing high-frequency local density fluctuations introduces slight noise for properties already well-defined by the rigid skeleton. Nonetheless, UniField maintains competitive second-best accuracy on these metrics, successfully integrating quantum priors without sacrificing geometric stability.
\begin{table*}[t]
  \caption{Ablation study evaluating the impact of fusion strategies and spatial distance bias across  ED5-OE and QM9-ED. The best results are \textbf{bolded} and the second-best are \underline{underlined}.}
  \label{tab:ablation_combined}
  \centering
  \renewcommand{\arraystretch}{1.2}
  \resizebox{\textwidth}{!}{
  \begin{tabular}{ccccccccccc}
    \toprule
    \multicolumn{11}{l}{\cellcolor{gray!10}\textit{\textbf{Dataset: ED5-OE (All MAE values in meV)}}} \\
    \midrule
    \textbf{ED Input} & \textbf{Spatial Fusion} & \textbf{Dist. Bias} & \textbf{HOMO-2} & \textbf{HOMO-1} & \textbf{HOMO-0} & \textbf{LUMO+0} & \textbf{LUMO+1} & \textbf{LUMO+2} & \textbf{LUMO+3} & \textbf{RMSE} \\
    \midrule
    $\times$ & $\times$ & $\times$ & 110.875 & 99.907 & \textbf{93.374} & \underline{102.990} & 120.049 & 126.845 & 127.839 & 165.653 \\
    $\checkmark$ & $\times$ & $\checkmark$ & \textbf{109.030} & 99.271 & 102.928 & 108.911 & 122.151 & 129.887 & 137.621 & 166.681 \\
    $\checkmark$ & $\checkmark$ & $\times$ & 110.366 & \underline{98.104} & 94.189 & 103.579 & \underline{116.727} & \underline{123.129} & \underline{125.023} & \underline{163.854} \\
    $\checkmark$ & $\checkmark$ & $\checkmark$ & \underline{109.434} & \textbf{96.427} & \underline{94.017} & \textbf{102.821} & \textbf{115.800} & \textbf{122.128} & \textbf{123.512} & \textbf{161.220} \\
    
    \midrule
    \midrule
    
    \multicolumn{11}{l}{\cellcolor{gray!10}\textit{\textbf{Dataset: QM9-ED}}} \\
    \midrule
    \multirow{2}{*}{\textbf{ED Input}} & \multirow{2}{*}{\textbf{Spatial Fusion}} & \multirow{2}{*}{\textbf{Dist. Bias}} & \textbf{$\mu$} & \textbf{HOMO} & \textbf{LUMO} & \textbf{$\Delta\varepsilon$} & \textbf{$\alpha$} & \textbf{$R^2$} & \textbf{ZPVE} & \textbf{$C_v$} \\
    & & & (mD) & (meV) & (meV) & (meV) & (mBohr$^3$) & (mBohr$^2$) & (meV) & (mcal/molK) \\
    \midrule
    $\times$ & $\times$ & $\times$ & 61.777 & \underline{38.062} & \underline{49.071} & \underline{59.575} & \textbf{254.779} & \textbf{4943.667} & \textbf{8.620} & \textbf{102.556} \\
    $\checkmark$ & $\times$ & $\checkmark$ & 61.094 & 41.206 & 53.454 & 64.890 & 275.117 & 5060.312 & 9.362 & 108.9245 \\
    $\checkmark$ & $\checkmark$ & $\times$ & \underline{42.633} & 43.566 & 54.181 & 66.355 & 287.343 & 5408.807 & 9.932 & 127.106 \\
    $\checkmark$ & $\checkmark$ & $\checkmark$ & \textbf{36.452} & \textbf{37.563} & \textbf{42.472} & \textbf{55.154} & \underline{256.416} & \underline{4969.021} & \underline{8.974} & \underline{107.778} \\
    \bottomrule
  \end{tabular}
  }
\end{table*}

\begin{table*}[htbp]
  \caption{Ablation study on QMugs-ED evaluating the impact of fusion strategies and spatial distance bias. All MAE values are reported in meV.}
  \label{tab:ablation_qmugsed}
  \centering
  \renewcommand{\arraystretch}{1.2}
  \resizebox{0.6\textwidth}{!}{
  \begin{tabular}{ccccccc}
    \toprule
    \textbf{ED Input} & \textbf{Spatial Fusion} & \textbf{Dist. Bias} & \textbf{HOMO} & \textbf{LUMO} & \textbf{$\Delta\varepsilon$} & \textbf{$\Delta E_f$} \\
    \midrule
    $\times$ & $\times$ & $\times$ & 71.939 & \underline{72.207} & \underline{95.685} & 567.167 \\
    $\checkmark$ & $\times$ & $\checkmark$ & \underline{70.867} & 74.044 & 97.103 & 3357.337 \\
    $\checkmark$ & $\checkmark$ & $\times$ & 71.931 & 74.223 & 95.727 & \textbf{510.254} \\
    $\checkmark$ & $\checkmark$ & $\checkmark$ & \textbf{62.492} & \textbf{64.118} & \textbf{84.592} & \underline{511.135} \\
    \bottomrule
  \end{tabular}
  }
\end{table*}

\subsection{Hyperparameter Analysis}
\label{subsec:hyperparameter}

To evaluate the architectural sensitivities of UniField, we conduct a hyperparameter analysis on the granularity of the spatial distance encoding during cross-modal fusion. 

The \texttt{num\_gaussians} parameter controls the resolution of the Radial Basis Function (RBF) encoding for electron-atom pairwise distances $r_{ij}$. As Table~\ref{tab:rbf_ablation} demonstrates, using too few kernels (8 or 16) restricts the expressiveness of the spatial bias, hindering the model's ability to capture subtle distance variations. Conversely, excessive kernels (64) introduce high-frequency noise and over-parameterization, slightly degrading the overall stability. A balance of 32 kernels achieves the optimal spatial granularity, effectively guiding cross-modal message passing to yield the lowest overall RMSE and superior accuracy across most frontier orbitals.

\begin{table}[htbp]
  \caption{Impact of the number of RBF Gaussian kernels on the ED5-OE benchmark. The $*$ indicates the default parameter used in our method.}
  \label{tab:rbf_ablation}
  \centering
  \footnotesize 
  \renewcommand{\arraystretch}{1.2}
  \resizebox{\columnwidth}{!}{
  \begin{tabular}{ccccccccc}
    \toprule
    \textbf{Num Gaussians} & \textbf{HOMO-2} & \textbf{HOMO-1} & \textbf{HOMO-0} & \textbf{LUMO+0} & \textbf{LUMO+1} & \textbf{LUMO+2} & \textbf{LUMO+3} & \textbf{RMSE} \\
    \midrule
    8 & 113.448 & 105.457 & 94.597 & 105.552 & 122.028 & 129.148 & 127.958 & 167.609 \\
    16 & \underline{108.632} & 102.582 & 94.828 & \underline{103.665} & 117.261 & \underline{122.401} & \underline{124.633} & 163.868 \\
    $^{\;\;}$ 32 $^{*}$ & 109.434 & \textbf{96.427} & \underline{94.017} & \textbf{102.821} & \underline{115.800} & \textbf{122.128} & \textbf{123.512} & \textbf{161.220} \\
    64 & \textbf{107.142} & \underline{99.401} & \textbf{93.218} & 104.638 & \textbf{115.500} & 124.189 & 125.383 & \underline{162.757} \\
    \bottomrule
  \end{tabular}
  }
\end{table}

\section{Conclusion}
\label{sec:conclusion}

In this work, we propose \textbf{UniField}, a molecular representation learning framework that natively fuses continuous electron density fields with discrete atomic graphs. Through a novel RBF-guided electron-to-atom interaction mechanism, UniField dynamically injects local quantum environments into SE(3)-equivariant atomic features. To enable systematic evaluation, we introduce the \textbf{UniField-ED benchmark}, derived from QM9 and QMugs, providing aligned graphs and high-fidelity density point clouds. Experiments across QM9-ED, QMugs-ED, and ED5-OE reveal that UniField significantly enhances property predictions, especially for frontier orbitals, while ablations validate our distance-guided cross-modal fusion. Ultimately, this work establishes electron density as a vital physical modality, advancing expressive molecular modeling for quantum chemistry and materials science.

{
\small
\bibliographystyle{unsrtnat}
\bibliography{references}
}




\appendix

\section{Background and Theoretical Foundations}
\label{app:background_theory}

\subsection{Electron Density as a Molecular Quantum Field}
\label{app:electron_density}

\subsubsection{Definition of Electron Density}
Electron density is a fundamental physical quantity in quantum chemistry. For an $N$-electron molecular system with many-electron wavefunction $\Psi(\mathbf{x}_1,\dots,\mathbf{x}_N)$, where $\mathbf{x}_i$ denotes both spatial and spin coordinates, the one-electron density $\rho(\mathbf{r})$ is defined as
\begin{equation}
    \rho(\mathbf{r})
    =
    N \int
    |\Psi(\mathbf{r}, s_1, \mathbf{x}_2,\dots,\mathbf{x}_N)|^2
    ds_1 d\mathbf{x}_2 \cdots d\mathbf{x}_N .
    \label{eq:ed_definition}
\end{equation}
Intuitively, $\rho(\mathbf{r})$ measures the expected electron population around spatial position $\mathbf{r}$. It satisfies the normalization condition
\begin{equation}
    \int \rho(\mathbf{r}) d\mathbf{r} = N,
    \label{eq:ed_normalization}
\end{equation}
where $N$ is the total number of electrons in the system.

Unlike atomistic graphs, which represent a molecule as a discrete set of nuclei and interatomic relations, electron density is a continuous scalar field distributed over three-dimensional space. It therefore contains information about electron delocalization, bonding regions, lone pairs, charge redistribution, and non-covalent interactions. These electronic effects are particularly important for quantum chemical properties such as frontier orbital energies, dipole moments, polarizability, and energy gaps.

\subsubsection{Electron Density and Molecular Properties}
The physical importance of electron density is formalized by Density Functional Theory (DFT). According to the Hohenberg--Kohn theorem, for a non-degenerate ground-state molecular system, the ground-state electron density uniquely determines the external potential up to an additive constant~\cite{hohenberg1964inhomogeneous}. Since the external potential determines the Hamiltonian of the system, the ground-state electron density in principle determines all ground-state properties of the molecule.

This theoretical result motivates the use of electron density as a molecular representation. While conventional 3D graph neural networks encode atomic numbers and nuclear coordinates, they do not explicitly represent the continuous electronic field induced by the molecular structure. In contrast, an electron-density-enhanced representation provides direct access to the microscopic electronic environment surrounding the atomic skeleton.

\subsection{Density Functional Theory}
\label{app:dft}

\subsubsection{From Many-Electron Wavefunctions to Electron Density}
The electronic structure of a molecule is governed by the time-independent Schr\"odinger equation:
\begin{equation}
    \hat{H} \Psi = E \Psi,
    \label{eq:schrodinger}
\end{equation}
where $\hat{H}$ is the Hamiltonian operator, $\Psi$ is the many-electron wavefunction, and $E$ is the total energy. Directly solving this equation for many-electron systems is computationally challenging because the wavefunction depends on the coordinates of all electrons. For an $N$-electron system, this leads to a high-dimensional many-body problem.

DFT reformulates the electronic structure problem in terms of the three-dimensional electron density $\rho(\mathbf{r})$ rather than the many-electron wavefunction. This greatly reduces the complexity of the representation while retaining the essential ground-state information of the system~\cite{parr1989density}.

\subsubsection{Kohn--Sham Formulation}
The Kohn--Sham formulation provides a practical computational framework for DFT~\cite{kohn1965self}. It introduces a set of auxiliary single-particle orbitals $\{\phi_i\}$ that reproduce the ground-state electron density:
\begin{equation}
    \rho(\mathbf{r}) = \sum_{i}^{\mathrm{occ}} |\phi_i(\mathbf{r})|^2 .
    \label{eq:ks_density}
\end{equation}
The Kohn--Sham orbitals are obtained by solving
\begin{equation}
    \left[
    -\frac{1}{2}\nabla^2 + V_{\mathrm{eff}}(\mathbf{r})
    \right]\phi_i(\mathbf{r})
    =
    \epsilon_i \phi_i(\mathbf{r}),
    \label{eq:ks_equation}
\end{equation}
where $V_{\mathrm{eff}}(\mathbf{r})$ is the effective single-particle potential:
\begin{equation}
    V_{\mathrm{eff}}(\mathbf{r})
    =
    V_{\mathrm{ext}}(\mathbf{r})
    +
    V_{\mathrm{H}}(\mathbf{r})
    +
    V_{\mathrm{xc}}(\mathbf{r}).
    \label{eq:effective_potential}
\end{equation}
Here, $V_{\mathrm{ext}}$ is the external potential induced by the nuclei, $V_{\mathrm{H}}$ is the Hartree potential describing classical electron-electron repulsion, and $V_{\mathrm{xc}}$ is the exchange-correlation potential that accounts for many-body quantum effects beyond the classical approximation.

This formulation shows that electron density is not merely an auxiliary descriptor, but a central variable through which molecular quantum properties are computed. It also provides the theoretical basis for constructing electron-density-enhanced molecular learning models.

\subsection{Electron Density for Molecular Representation Learning}
\label{app:ed_representation}

\subsubsection{Limitations of Discrete Atomic Graphs}
Most 3D molecular learning models represent a molecule as a set of atoms with coordinates and pairwise geometric relations. Such representations are efficient and naturally compatible with graph neural networks, but they primarily describe the nuclear skeleton. The surrounding electronic distribution is only implicitly inferred from atom types and geometry.

This design can be restrictive for properties that depend strongly on electronic structure. For example, frontier orbital energies and HOMO--LUMO gaps are closely related to electron delocalization and the spatial distribution of molecular orbitals. Non-covalent interactions and charge redistribution are also difficult to fully characterize using only discrete atomic nodes and interatomic distances. These observations motivate the explicit incorporation of electron density into molecular representations.

\subsubsection{Electron Density as a Point Cloud}
A direct way to represent electron density is to discretize $\rho(\mathbf{r})$ on a dense 3D voxel grid. However, dense voxelization suffers from cubic memory and computational complexity with respect to spatial resolution. This makes high-resolution electron density modeling expensive, especially for large molecules.

In UniField-ED, the continuous density field is instead represented as a point cloud:
\begin{equation}
    \mathcal{P}
    =
    \{(\mathbf{p}_j, e_j)\}_{j=1}^{M},
    \label{eq:ed_point_cloud_app}
\end{equation}
where $\mathbf{p}_j \in \mathbb{R}^3$ is the coordinate of the sampled point and $e_j = \rho(\mathbf{p}_j)$ is the corresponding density value. This representation preserves the native 3D geometry of the electron density field while avoiding the cubic cost of voxel grids. It also allows the model to focus computation on physically meaningful regions where the electron density is non-negligible.

\subsection{RBF-Guided Electron-to-Atom Fusion}
\label{app:rbf_fusion_theory}

\subsubsection{Need for Spatially Grounded Fusion}
Although electron density provides rich physical information, simply adding it as a separate global feature is insufficient. The electron density point cloud and the atomic graph are defined on different supports: the former consists of sampled field points in continuous space, while the latter consists of discrete atomic centers. Therefore, effective fusion requires explicit spatial correspondence between density points and atoms.

UniField addresses this by using electron-to-atom interaction. For each atom $i$ located at $\mathbf{r}_i$, nearby electron density points are selected according to a cutoff radius $R_c$:
\begin{equation}
    \mathcal{N}(i)
    =
    \{j \mid \|\mathbf{p}_j - \mathbf{r}_i\|_2 \le R_c\}.
    \label{eq:local_ed_neighborhood_app}
\end{equation}
This local neighborhood defines the electronic environment around each atomic anchor.

\subsubsection{Radial Basis Distance Bias}
To encode the relative distance between atom $i$ and electron density point $j$, UniField expands the Euclidean distance
\begin{equation}
    d_{ij} = \|\mathbf{p}_j - \mathbf{r}_i\|_2
\end{equation}
using Gaussian radial basis functions:
\begin{equation}
    e_{ij}^{(k)}
    =
    \exp\left(
    -\gamma (d_{ij} - \mu_k)^2
    \right),
    \quad k = 1,\dots,K.
    \label{eq:rbf_app}
\end{equation}
The RBF embedding provides a smooth and continuous encoding of atom--density distances. Compared with using raw distances, RBF features offer a more expressive geometric basis and have been widely used in molecular neural networks to encode interatomic distances~\cite{schutt2017schnet}.

In UniField, the distance embedding is converted into a scalar bias and added to the attention score:
\begin{equation}
    s_{ij}
    =
    \frac{\mathbf{q}_i^\top \mathbf{k}_j}{\sqrt{d}}
    +
    \mathrm{MLP}_{\mathrm{dist}}(\mathbf{e}_{ij}).
    \label{eq:rbf_attention_app}
\end{equation}
This design encourages the model to consider both semantic feature compatibility and physical spatial proximity. As a result, the fused atomic representation can absorb local electronic information from the surrounding density field in a distance-aware manner.

\subsection{Symmetry Considerations}
\label{app:symmetry}

Molecular property prediction should be invariant to global translations, rotations, atom indexing, and the ordering of sampled density points. The RBF-guided fusion mechanism satisfies these requirements at the scalar-feature level because the distance term depends only on $\|\mathbf{p}_j-\mathbf{r}_i\|_2$, which is invariant under rigid transformations.

Consider a global transformation
\begin{equation}
    \mathbf{r}_i' = Q\mathbf{r}_i + \mathbf{t},
    \quad
    \mathbf{p}_j' = Q\mathbf{p}_j + \mathbf{t},
    \label{eq:rigid_transform_app}
\end{equation}
where $Q$ is an orthogonal rotation matrix and $\mathbf{t}$ is a translation vector. Then,
\begin{equation}
    \|\mathbf{p}_j' - \mathbf{r}_i'\|_2
    =
    \|Q(\mathbf{p}_j-\mathbf{r}_i)\|_2
    =
    \|\mathbf{p}_j-\mathbf{r}_i\|_2.
    \label{eq:distance_invariance_app}
\end{equation}
Thus, the RBF distance bias remains unchanged under global rotations and translations. In addition, the aggregation over electron density points is permutation-invariant with respect to their ordering, and the final molecular readout uses permutation-invariant pooling over atoms. Therefore, UniField preserves the essential symmetry requirements for scalar molecular property prediction while incorporating continuous electron density information.

\section{Dataset Construction and Benchmark Details}
\label{app:dataset_details}

This section provides detailed information on the construction of the UniField-ED benchmark, including source datasets, quantum chemistry calculation settings, electron density generation, point-cloud sampling, data schemas, and target-property processing. The goal of UniField-ED is to provide a standardized multimodal benchmark in which each molecule is represented by both a discrete atomic graph and a spatially aligned continuous electron density field.

\subsection{Overview of UniField-ED}
\label{app:dataset_overview}

UniField-ED consists of two electron-density-enhanced molecular datasets: QM9-ED and QMugs-ED. QM9-ED is built from the complete QM9 database~\cite{ramakrishnan2014qm9}, covering small organic molecules with up to 15 atoms. QMugs-ED is constructed from a curated subset of QMugs~\cite{isert2022qmugs}, containing larger and more structurally diverse drug-like conformations. Together, the two datasets provide complementary evaluation regimes: QM9-ED focuses on relatively small molecules with localized electronic structures, while QMugs-ED evaluates whether electron-density-enhanced representations remain effective for larger molecules with more complex electronic environments.

For each molecule, the atomic structure is retained as the geometric modality $\mathcal{G}$, and the corresponding electron density field is processed as the continuous physical-field modality $\mathcal{P}$. Both modalities are stored in a unified format, enabling controlled comparison between geometry-only models, electron-density-only models, and multimodal fusion models.

\subsection{Quantum Chemistry Calculations}
\label{app:qc_calculations}

\textbf{QM9-ED.}
For QM9-ED, the molecular structures and quantum chemical annotations are derived from the QM9 database. The original QM9 molecules are associated with DFT calculations at the B3LYP/6-31G(2df,p) level, with several thermochemical quantities reported using G4MP2-related annotations. To construct the electron density modality, we use the Psi4 quantum chemistry package~\cite{smith2020psi4} to compute electron density fields for the QM9 molecules under the corresponding DFT setting. The resulting density field is spatially aligned with the molecular coordinates and subsequently converted into a point-cloud representation.

\textbf{QMugs-ED.}
For QMugs-ED, we curate a subset of $50{,}000$ drug-like conformations from QMugs. QMugs provides molecular conformations and quantum chemical properties generated from semi-empirical and DFT calculations. In our benchmark construction, the QMugs-derived conformations are used to generate aligned molecular structures and electron density fields, with target properties including formation energy, exchange-correlation energy, HOMO energy, LUMO energy, and HOMO--LUMO gap. The DFT-level annotations follow the $\omega$B97X-D/def2-SVP setting, while GFN2-xTB information is retained as part of the original QMugs computational pipeline. This setting provides a more challenging benchmark than QM9-ED due to the increased molecular size and structural diversity.

\subsection{Electron Density Field Generation}
\label{app:ed_generation}

For a molecule with atomic coordinates $\{\mathbf{r}_i\}_{i=1}^{N}$ and atom types $\{z_i\}_{i=1}^{N}$, the electron density is a continuous scalar field
\begin{equation}
    \rho(\mathbf{r}): \mathbb{R}^{3} \rightarrow \mathbb{R},
\end{equation}
where $\rho(\mathbf{r})$ denotes the electron density value at spatial position $\mathbf{r}$. The raw electron density field produced by quantum chemistry calculation is defined on a three-dimensional spatial grid. Directly using the full grid is computationally expensive because dense voxelization scales cubically with spatial resolution. Therefore, UniField-ED converts the continuous field into a compact point-cloud representation.

The density generation and formatting pipeline follows three principles. First, the generated density field must remain spatially aligned with the corresponding atomic coordinates. Second, the representation should preserve high-density regions that are chemically meaningful, such as bonding regions, lone-pair regions, and delocalized electron clouds. Third, the resulting representation should have a fixed size to support efficient mini-batch training.

\subsection{Coordinate Alignment and Unit Consistency}
\label{app:coordinate_alignment}

A critical requirement for electron-density-enhanced molecular learning is that atomic coordinates and electron density points are represented in the same coordinate frame. In UniField-ED, the electron density point cloud and the atomic graph are stored as aligned 3D structures. Each electron density point has a coordinate $\mathbf{p}_j$ that is directly comparable to atomic coordinates $\mathbf{r}_i$ during electron-to-atom fusion.

When raw quantum chemistry outputs and molecular coordinates use different length units, the coordinates are converted into a consistent unit before model input. In particular, if electron density grid coordinates are produced in Bohr while atomic coordinates are represented in Angstrom, the following conversion is applied:
\begin{equation}
    1~\mathrm{Bohr} = 0.529177~\mathrm{\AA}.
\end{equation}
This coordinate consistency is essential for distance-based operations such as radius neighborhood construction and RBF-guided electron-to-atom fusion.

\subsection{Point-Cloud Sampling Pipeline}
\label{app:point_sampling_pipeline}

Given a raw electron density grid, UniField-ED constructs a fixed-size electron density point cloud using a two-stage sampling pipeline.

\textbf{Density thresholding.}
The raw grid contains many points in low-density vacuum regions that contribute little chemical information but greatly increase computational cost. Therefore, grid points whose density values are below a predefined threshold $\tau$ are removed:
\begin{equation}
    \mathcal{P}_{\mathrm{valid}}
    =
    \left\{
    (\mathbf{p}_j, e_j)
    \mid
    e_j = \rho(\mathbf{p}_j) \geq \tau
    \right\}.
\end{equation}
This step preserves chemically informative density regions while removing redundant low-density space.

\textbf{Farthest point sampling.}
After thresholding, the number of remaining density points varies across molecules. To obtain a fixed-size representation, farthest point sampling (FPS) is applied to select $M=1024$ points:
\begin{equation}
    \mathcal{P}
    =
    \{(\mathbf{p}_j, e_j)\}_{j=1}^{1024}.
\end{equation}
FPS encourages spatial coverage over the valid density region and avoids over-concentration in highly dense local areas. As a result, each molecule is represented by an electron density point cloud of shape
\begin{equation}
    \texttt{ed} \in \mathbb{R}^{1024 \times 4},
\end{equation}
where the four channels correspond to $(x,y,z,e)$.

The atomic graph modality is stored separately as
\begin{equation}
    \texttt{atom\_coords} \in \mathbb{R}^{N \times 3},
    \quad
    \texttt{atom\_types} \in \mathbb{R}^{N},
\end{equation}
where $N$ is the number of atoms in the molecule. The number of atoms is molecule-dependent. In the processed datasets, inspected samples from QM9-ED and QMugs-ED contain atomic coordinate arrays of shape $(15,3)$ and $(80,3)$, respectively.

\subsection{Dataset Splits}
\label{app:dataset_splits}

The processed UniField-ED datasets are stored as Python dictionaries with three top-level keys:
\begin{equation}
    \{\texttt{train}, \texttt{valid}, \texttt{test}\}.
\end{equation}
The same split is used across all compared methods to ensure fair evaluation.

\textbf{QM9-ED split.}
QM9-ED contains $133{,}885$ molecules. The train/validation/test split is
\begin{equation}
    107{,}108 / 13{,}388 / 13{,}389.
\end{equation}

\textbf{QMugs-ED split.}
QMugs-ED contains $50{,}000$ molecules. The train/validation/test split is
\begin{equation}
    40{,}000 / 5{,}000 / 5{,}000.
\end{equation}

\subsection{Stored Fields in QM9-ED}
\label{app:qm9_schema}

Each QM9-ED sample contains both molecular structural information and quantum chemical target properties. The stored fields are:
\begin{equation}
\begin{aligned}
\{&
\texttt{ed},\
\texttt{atom\_coords},\
\texttt{atom\_types},\
\texttt{energy},\
\texttt{mu\_mD},\
\texttt{alpha\_mBohr3},\\
&
\texttt{homo\_meV},\
\texttt{lumo\_meV},\
\texttt{gap\_meV},\
\texttt{r2\_mBohr2},\
\texttt{zpve\_meV},\\
&
\texttt{U0\_meV},\
\texttt{U298\_meV},\
\texttt{H298\_meV},\
\texttt{G298\_meV},\
\texttt{Cv\_mcal}
\}.
\end{aligned}
\end{equation}

The main QM9-ED benchmark targets used in our experiments include dipole moment $\mu$, polarizability $\alpha$, HOMO energy, LUMO energy, HOMO--LUMO gap, electronic spatial extent $R^2$, zero-point vibrational energy, and heat capacity. The property units are summarized as follows:
\begin{equation}
\begin{aligned}
&\mu: \mathrm{mD}, \quad
\alpha: \mathrm{mBohr}^{3}, \quad
R^2: \mathrm{mBohr}^{2}, \\
&\mathrm{HOMO}, \mathrm{LUMO}, \Delta\varepsilon, \mathrm{ZPVE}: \mathrm{meV}, \quad
C_v: \mathrm{mcal/molK}.
\end{aligned}
\end{equation}

\subsection{Stored Fields in QMugs-ED}
\label{app:qmugs_schema}

Each QMugs-ED sample contains the following fields:
\begin{equation}
\begin{aligned}
\{&
\texttt{id},\
\texttt{ed},\
\texttt{atom\_coords},\
\texttt{atom\_types},\
\texttt{energy},\\
&
\texttt{formation\_energy\_meV},\
\texttt{xc\_energy\_meV},\
\texttt{homo\_energy\_meV},\\
&
\texttt{lumo\_energy\_meV},\
\texttt{homo\_lumo\_gap\_meV}
\}.
\end{aligned}
\end{equation}

The main QMugs-ED benchmark targets used in this work include formation energy, HOMO energy, LUMO energy, and HOMO--LUMO gap. All reported target values for these tasks are represented in meV.

\subsection{Target Unit Conversion}
\label{app:unit_conversion}

Energy-related quantities are converted to meV when necessary to maintain a consistent evaluation scale across datasets:
\begin{equation}
    1~\mathrm{Hartree}
    =
    27211.386245988532898~\mathrm{meV}.
\end{equation}
This conversion is applied to frontier orbital energies, HOMO--LUMO gaps, and thermodynamic energy-related quantities when the original values are provided in Hartree. Using a unified energy unit also makes the reported MAE values directly comparable across QM9-ED, QMugs-ED, and ED5-OE.

\subsection{Benchmark Summary}
\label{app:benchmark_summary}

Table~\ref{tab:app_dataset_schema} summarizes the processed dataset statistics and the main stored target properties.

\begin{table}[htbp]
  \caption{Detailed schema of the UniField-ED benchmark. Both datasets contain aligned atomic graphs and electron density point clouds of shape $(1024,4)$.}
  \label{tab:app_dataset_schema}
  \centering
  \renewcommand{\arraystretch}{1.25}
  \resizebox{\textwidth}{!}{
  \begin{tabular}{lcccc}
    \toprule
    \textbf{Dataset} & \textbf{Total} & \textbf{Train/Valid/Test} & \textbf{ED Shape} & \textbf{Main Stored Targets} \\
    \midrule
    QM9-ED
    & $133{,}885$
    & $107{,}108 / 13{,}388 / 13{,}389$
    & $(1024,4)$
    & $\mu$, $\alpha$, HOMO, LUMO, gap, $R^2$, ZPVE, $C_v$ \\
    \midrule
    QMugs-ED
    & $50{,}000$
    & $40{,}000 / 5{,}000 / 5{,}000$
    & $(1024,4)$
    & $E_{\mathrm{form}}$, HOMO, LUMO, gap \\
    \bottomrule
  \end{tabular}
  }
\end{table}

Overall, UniField-ED provides a consistent multimodal benchmark for electron-density-enhanced molecular learning. Each molecule contains an aligned atomic graph, a high-fidelity electron density point cloud, and quantum chemical target properties. This design allows controlled evaluation of whether continuous electronic fields can provide complementary information beyond conventional geometry-only molecular representations.

\subsection{Computational Cost of Benchmark Construction}
\label{app:computational_cost}

The construction of the UniField-ED benchmark involves extensive quantum chemistry computations, and the release of this processed resource represents a significant reduction in redundant calculation for the AI-for-Science community. 

Using the Psi4 quantum chemistry package~\cite{smith2020psi4}, we executed DFT self-consistent field (SCF) calculations and density field generation for a total of \textbf{183,885} molecules (133,885 from QM9 and 50,000 from QMugs). Each individual calculation was allocated 16 CPU threads and 4 GB of memory. Given an average processing time of approximately \textbf{3 minutes} per molecule under this configuration, the generation of the entire benchmark required approximately \textbf{147,100 CPU hours} on a high-performance computing (HPC) cluster. 

Specifically, the QMugs-ED subset alone accounted for 40,000 CPU hours, while the expansive QM9-ED dataset required over 107,000 CPU hours. Beyond the processing time, the intermediate voxelized density files (Cube format) generated during this pipeline occupied several terabytes of temporary storage before being distilled into our optimized, fixed-size $(1024, 4)$ point-cloud modality. By providing these natively aligned and high-fidelity representations, we effectively bypass these prohibitive computational and storage barriers, facilitating the efficient development and benchmarking of future electron-density-enhanced molecular models.

\section{Training and Implementation Details}
\label{app:training_details}

This section provides additional implementation details for training UniField and all compared variants. To ensure fair comparison, all models are trained under the same dataset split and evaluation protocol for each benchmark. Unless otherwise specified, the random seed is fixed to $2025$ for all experiments.

\subsection{Training Configuration}
\label{app:training_config}

The training hyperparameters used for different benchmarks are summarized in Table~\ref{tab:app_training_config}. ED5-OE is trained with a smaller batch size due to the larger multimodal input and the larger number of orbital prediction targets. QM9-ED and QMugs-ED use the same training configuration.

\begin{table}[htbp]
  \caption{Training hyperparameters used in UniField experiments.}
  \label{tab:app_training_config}
  \centering
  \renewcommand{\arraystretch}{1.2}
  \resizebox{\textwidth}{!}{
  \begin{tabular}{lccccccc}
    \toprule
    \textbf{Dataset} & \textbf{Seed} & \textbf{Epochs} & \textbf{Patience} & \textbf{Batch Size} & \textbf{Learning Rate} & \textbf{Weight Decay} & \textbf{Grad. Clip} \\
    \midrule
    ED5-OE 
    & $2025$ & $300$ & $50$ & $16$ & $1.0\times10^{-4}$ & $1.0\times10^{-4}$ & $1.0$ \\
    QM9-ED 
    & $2025$ & $200$ & $50$ & $48$ & $1.0\times10^{-4}$ & $1.0\times10^{-4}$ & $1.0$ \\
    QMugs-ED 
    & $2025$ & $200$ & $50$ & $48$ & $1.0\times10^{-4}$ & $1.0\times10^{-4}$ & $1.0$ \\
    \bottomrule
  \end{tabular}
  }
\end{table}

\subsection{Optimization}
\label{app:optimization}

All models are optimized end-to-end using the Adam optimizer with a learning rate of $1.0\times10^{-4}$ and a weight decay of $1.0\times10^{-4}$. A cosine annealing learning-rate scheduler is applied over the maximum number of training epochs. To stabilize training, gradient norm clipping with threshold $1.0$ is used. Early stopping is applied with a patience of $50$ epochs according to the validation RMSE.

For multi-task prediction, the model output dimension is determined by the number of target properties defined in the dataset configuration. Given $T$ target properties, the model predicts a vector $\hat{\mathbf{y}}\in\mathbb{R}^{T}$. The training objective is the mean squared error over Z-score normalized targets:
\begin{equation}
    \mathcal{L}
    =
    \frac{1}{T}
    \sum_{t=1}^{T}
    \left(
    \hat{y}_{t}^{\,\mathrm{norm}}
    -
    y_{t}^{\,\mathrm{norm}}
    \right)^2 .
\end{equation}
The normalization statistics are computed only from the training split:
\begin{equation}
    y_{t}^{\,\mathrm{norm}}
    =
    \frac{y_t - \mu_t}{\sigma_t},
\end{equation}
where $\mu_t$ and $\sigma_t$ denote the mean and standard deviation of the $t$-th target property in the training set. During evaluation, predictions are inverse-transformed back to the original physical units before computing MAE and RMSE:
\begin{equation}
    \hat{y}_t
    =
    \hat{y}_{t}^{\,\mathrm{norm}}\sigma_t + \mu_t .
\end{equation}

\subsection{Data Loading and Reproducibility}
\label{app:data_loading}

The training pipeline loads three configuration files for each experiment: a training configuration, a data configuration, and a model configuration. These files define the training hyperparameters, dataset modality, target-property list, and model architecture. For reproducibility, all configuration files are automatically copied into the corresponding experiment log directory before training starts.

Mini-batches are constructed using dataset-specific collation functions to support different input modalities, including atomic graphs, electron density point clouds, and multimodal graph-field inputs. For point-cloud and multimodal datasets, we use $8$ data-loading workers. For atomistic graph-only experiments, the number of workers is set to $0$ for compatibility.

The best checkpoint is selected according to the lowest validation RMSE. Each saved checkpoint contains the model parameters, readout parameters, optimizer state, target normalization statistics, best validation RMSE, and the list of target properties. This ensures that evaluation can be performed using the same normalization and target ordering as training.

\subsection{Evaluation Metrics}
\label{app:evaluation_metrics}

Following standard practice in quantum property prediction, all reported metrics are computed in the original physical units after inverse normalization. Mean Absolute Error (MAE) is used as the primary metric for individual target properties:
\begin{equation}
    \mathrm{MAE}
    =
    \frac{1}{n}
    \sum_{i=1}^{n}
    \left|
    \hat{y}_i - y_i
    \right| .
\end{equation}
For multi-target orbital prediction on ED5-OE, we additionally report the average RMSE across target properties:
\begin{equation}
    \mathrm{RMSE}
    =
    \sqrt{
    \frac{1}{nT}
    \sum_{i=1}^{n}
    \sum_{t=1}^{T}
    \left(
    \hat{y}_{i,t} - y_{i,t}
    \right)^2
    } .
\end{equation}
Lower MAE and RMSE indicate better predictive accuracy.

\section{Algorithm Pseudocode of UniField Forward Pass}
\label{app:pseudocode}

Algorithm~\ref{alg:unifield} provides the detailed computational workflow of the proposed UniField representation framework, illustrating the processing of geometric and physical modalities, the spatially-grounded cross-modal interaction, and the multi-task property prediction.

\begin{algorithm}[htbp]
\caption{Forward Pass of  UniField }
\label{alg:unifield}
\textbf{Input}: Atomic graph $\mathcal{G} = \{X_{atom}, P_{atom}\}$ with $N_g$ nodes, ED point cloud $\mathcal{P} = \{X_{cloud}, P_{cloud}\}$ with $N_p$ points, interaction radius $r$, maximum neighbors $K$, hidden dimension $d$, multi-task count $T$.\\
\textbf{Output}: Multi-task property predictions $Y \in \mathbb{R}^{T}$
\begin{algorithmic}[1]
\STATE \textit{// Step 1: Modality-Specific Feature Extraction}
\STATE $H_g \leftarrow \text{Equiformer}(X_{atom}, P_{atom})$ \hfill \textit{Extract SE(3)-equivariant geometric tokens}
\STATE $H_p \leftarrow \text{PTv3}(X_{cloud}, P_{cloud})$ \hfill \textit{Extract continuous field tokens}

\STATE \textit{// Step 2: Spatially-Grounded Cross-Modal Interaction}
\STATE $Q \leftarrow H_g W_q, \quad K \leftarrow H_p W_k, \quad V \leftarrow H_p W_v$ \hfill \textit{Linear projections}
\STATE $\mathcal{E}_{int} \leftarrow \text{RadiusSearch}(P_{cloud}, P_{atom}, r, K)$ \hfill \textit{Find interacting edges $(i, j)$ where $j \in \mathcal{N}(i)$}

\FOR{each atom $i \in \{1, \dots, N_g\}$}
    \FOR{each neighboring field point $j \in \mathcal{N}(i)$}
        \STATE $D_{ij} \leftarrow \| P_{cloud}^{(j)} - P_{atom}^{(i)} \|_2$ \hfill \textit{Calculate Euclidean distance}
        \STATE $\text{RBF}_{ij} \leftarrow \text{GaussianSmearing}(D_{ij})$ \hfill \textit{Expand via Gaussian kernels}
        \STATE $B_{ij} \leftarrow \text{MLP}_{dist}(\text{RBF}_{ij})$ \hfill \textit{Compute RBF-guided distance bias}
        \STATE $S_{ij} \leftarrow \frac{Q_i K_j^T}{\sqrt{d}} + B_{ij}$ \hfill \textit{Calculate biased attention score}
    \ENDFOR
    \STATE $\alpha_{ij} \leftarrow \text{Softmax}_j(S_{ij})$ \hfill \textit{Normalize across neighbors $j$}
    \STATE $M_i \leftarrow \sum_{j \in \mathcal{N}(i)} \alpha_{ij} V_j$ \hfill \textit{Aggregate field messages to atom}
    \STATE $H_{g, i}^{new} \leftarrow \text{LayerNorm}\Big(H_{g, i} + \text{Dropout}\big(\text{MLP}_{out}(M_i)\big)\Big)$ \hfill \textit{Update atom token}
\ENDFOR

\STATE \textit{// Step 3: Global Aggregation and Readout}
\STATE $Z \leftarrow \text{MLP}_{proj}(H_g^{new})$ \hfill \textit{Project refined representations}
\STATE $Z_{sum} \leftarrow \sum_{i=1}^{N_g} Z_i, \quad Z_{mean} \leftarrow \frac{1}{N_g}\sum_{i=1}^{N_g} Z_i, \quad Z_{max} \leftarrow \max_{i} Z_i$
\STATE $Z_{global} \leftarrow \text{Concat}(Z_{sum}, Z_{mean}, Z_{max})$ \hfill \textit{Multi-pooling aggregation}

\STATE \textit{// Step 4: Multi-Task Independent Prediction}
\FOR{$t \in \{1, \dots, T\}$}
    \STATE $Y_t \leftarrow \text{Head}_t(Z_{global})$ \hfill \textit{Task-specific MLP prediction}
\ENDFOR

\STATE \textbf{return} $Y$
\end{algorithmic}
\end{algorithm}

\section{Limitations}
\label{app:limitations}

Although UniField shows consistent benefits from incorporating electron density into molecular representations, several limitations remain. First, constructing high-fidelity electron density fields requires offline quantum chemistry calculations with Psi4 and DFT-level settings, introducing additional preprocessing cost compared with geometry-only benchmarks. Second, the multimodal design increases training complexity and GPU memory usage, since UniField processes both atomic graphs and electron density point clouds and performs electron-to-atom fusion between them. Although the radius-restricted interaction avoids global atom-density attention, larger molecules or higher-resolution density point clouds may still require more computational resources and smaller batch sizes. Finally, the quality and applicability of UniField depend on the underlying density calculation, grid resolution, sampling strategy, and benchmark coverage. Current evaluations mainly focus on QM9, QMugs, and ED5-OE, and extending the framework to broader chemical systems such as periodic materials, transition-metal complexes, charged molecules, and biomolecules remains future work.


\end{document}